\documentclass[12pt]{article}

\newcommand{\be}{\begin{equation}}
\newcommand{\ee}{\end{equation}}
\newcommand{\ba}{\begin{eqnarray}}
\newcommand{\ea}{\end{eqnarray}}
\newcommand{\dif}{\mathrm{d}}

\usepackage{graphicx}
\usepackage{amssymb}
\usepackage{amsmath}
\usepackage[T1]{fontenc} 
\usepackage[ansinew]{inputenc} 
\usepackage[nosort]{cite}
\newcommand{\inbar}{\vrule height1.57ex width.4pt depth0pt}
\newcommand{\SW}{\relax{\hbox{$\ \inbar\kern-.285em{\rm S}$}}}
\begin{document}
\thispagestyle{empty}
\begin{center}

\null \vskip-1truecm \vskip2truecm

{\Large{\bf \textsf{Isoperimetric Inequalities and Magnetic Fields at CERN}}}

{\large{\bf \textsf{}}}

{\large{\bf \textsf{}}}

\vskip1truecm

{\large \textsf{Brett McInnes}}

\vskip1truecm

\textsf{\\  National
  University of Singapore}

\textsf{email: matmcinn@nus.edu.sg}\\

\end{center}
\vskip1truecm \centerline{\textsf{ABSTRACT}} \baselineskip=15pt
\medskip

We discuss the generalization of the classical isoperimetric inequality to asymptotically hyperbolic Riemannian manifolds. It has been discovered that the AdS/CFT correspondence
in string theory requires that such an inequality hold in order to be internally consistent. In a particular application, to the systems formed in collisions of heavy ions
in particle colliders, we show how to formulate this inequality in terms of measurable physical quantities, the magnetic field and the temperature. Experiments under way
at CERN in Geneva can thus be said to be testing an isoperimetric inequality.

\newpage

\newpage
\addtocounter{section}{1}
\section* {\large{\textsf{1. The Isoperimetric Inequality and its Generalizations}}}
The notion that there must be some \emph{universal} inequality relating the volume of a finite region in space, and (a suitable power of) the area of the boundary of that region, has intrigued mathematicians since antiquity \cite{kn:iso}: one speaks of \emph{isoperimetric inequalities}. One of the most important developments in recent theoretical physics has been the realization, beginning with Juan Maldacena's celebrated 1997 work\footnote{This paper has now been cited over 11000 times.} \cite{kn:malda}, that there is also a certain (extremely deep) relation between the \emph{physics} in certain regions of space, and the physics on the boundary of that region. The two relations, one in geometry, the other in (sometimes surprisingly concrete) physics, are themselves related: and this is the theme of the present work.

In the simplest case, the isoperimetric inequality states the following. Let $L$ be the length of a closed curve in the plane, and $A$ be the area it encloses. Then
\begin{equation}\label{A}
A \;\leq \;{1\over 4\pi}L^2,
\end{equation}
with equality holding if and only if the curve is a circle. A useful way of re-stating this is as follows. Consider a circle of circumference $L$, containing an area $A$; then $A \;= \;{1\over 4\pi}L^2$. Now continuously distort the circle in such a way that the circumference remains equal to $L$: the result is always to \emph{decrease} the area below ${1\over 4\pi}L^2$.

As with so many familiar facts in mathematics, it is worth while to stop for a moment and reflect on how extraordinary this simple relation really is. What it is saying is that, \emph{by knowing something about the boundary, one immediately knows something about what is happening deep inside}, perhaps very far from the boundary: one cannot have a vast area lurking inside a small boundary. That is far from obvious, and imaginative people \cite{kn:tardis} have often entertained other possibilities.

Indeed, it is easy to see that, if we allow arbitrary geometries, then no isoperimetric inequality is possible: we just have to imagine that the interior is made of (mathematical) rubber, which we can distort to any size without changing the length of the boundary. One obtains the inequality (\ref{A}) only by using the fact that the geometry is planar. This seemingly trivial constraint on the geometry must be deeper than it looks: the internal consistency of planar geometry somehow implies that only those pairs ($A,\,L$) which satisfy (\ref{A}) are possible.

In other, non-planar geometries, analogues of the isoperimetric inequality can still exist, but they often take quite different forms to (\ref{A}). For example, consider a four-dimensional space with Cartesian topology and coordinates $(r,t,x,y)$, in which distances are measured\footnote{Readers who are not familiar with the details of Riemannian geometry are invited to interpret this formula in the obvious way: that is, $dr$ represents a ``small'' change in $r$, and so on, and the formula itself is a straightforward generalization of Pythagoras' theorem.} according to the metric
\begin{equation}\label{B}
g\;=\;dr^2\;+\;e^{2r/\rho}\,\left(dt^2\;+\;dx^2\;+\;dy^2\right),
\end{equation}
where $\rho$ is a certain positive constant. Now consider a finite domain $0\;\leq r \leq R,\;0\;\leq t \leq T,\;0\;\leq x \leq X,\;0\;\leq y \leq Y,$ and consider the subspace defined by $r = R$: think of it as the three-dimensional boundary of this four-dimensional domain. (That is, ignore the ``sides'' and focus on the outer surface.) The ``volume'' of this boundary (it is usually called the ``area'' of this ``surface'', not to be confused with the two-dimensional area we discussed earlier), $A_R$, is clearly $A_R = TXYe^{3R/\rho}.$ Since it measures the size of the boundary, this will be the analogue of the circumference $L$ in (\ref{A}). Next, compute the volume $V_R$ enclosed by this surface, measured outwards from $r = - \infty$: it is just ${\rho \over 3}TXYe^{3R/\rho}$, and this of course is the analogue of the area $A$ in (\ref{A}). So we have
\begin{equation}\label{C}
A_R\;=\;{3\over \rho}\,V_R.
\end{equation}

Now in fact, the space with which we are dealing here is just\footnote{Technical point: we are using possibly unfamiliar coordinates, which in fact do not cover all of hyperbolic space.} four-dimensional \emph{hyperbolic space}, the space of constant negative curvature $-1/\rho^2$. This space is maximally symmetric (it is the negatively curved analogue of the sphere), and so the situation we have been considering is analogous to the case of equality in (\ref{A}): for one has equality only in the case of a maximally symmetric region in the plane, namely the circle. That is, equation (\ref{C}) is a special case of an ``isoperimetric inequality'', with four-dimensional hyperbolic space playing the role of the circular disc.

If we pursue this analogy, then we should expect that the isoperimetric inequality itself should take the form
\begin{equation}\label{D}
A_R\;-\;{3\over \rho}\,V_R\;\geq\;0
\end{equation}
for all $R$ and for some class of distortions of four-dimensional hyperbolic space ---$\,$ recall that we interpreted (\ref{A}) as stating that equality becomes a strict inequality when the circular disc is distorted.

Notice that (\ref{D}) differs, in one important way, from (\ref{A}): it does not compare the volume with a power of the area, but rather with the area itself. An inequality like (\ref{D}) could not possibly hold everywhere in Euclidean space, because, as the surface becomes larger, the volume will eventually grow much more quickly than the area. Hyperbolic space is different: as we saw above, the area and the volume \emph{both} grow exponentially as the surface increases in size. One can in fact see this in a qualitative way by means of a close study of M.C. Escher's ``Circle Limit'' illustrations: see for example ``Circle Limit IV'' \cite{kn:escher}.

The question now is this: for which ``class of distortions'' of hyperbolic space does (\ref{D}) hold? (Of course it will not hold if we allow arbitrary distortions, just as in the case of the circle.) As might be expected, this is a very difficult question to answer, but some interesting results have been established.

Even to state these results in full rigour would take us too far afield, so I will attempt to state them in a non-rigorous way, leaving the interested reader to consult the references below for full definitions.

First, we need the concept of an \emph{asymptotically hyperbolic} manifold: (very) roughly speaking, this is a space which resembles hyperbolic space more and more closely ``towards infinity''. (Again, imagine taking ``Circle Limit IV'' and distorting the central region, while leaving the region ``near'' the boundary intact.)

Next, we need the concept of an \emph{Einstein manifold}. The familiar concept of the curvature of a surface, usually called the Gaussian curvature, can be generalized to higher dimensions. One begins by recognising that while a smooth two-dimensional surface has a well-defined tangent plane at each point, higher dimensional tangent spaces can be ``sliced'' in many different directions at a given point. The result is that a given manifold can be curved in different ways even at a given point, depending on the direction of the ``slicing''. An asymptotically hyperbolic Einstein manifold is one in which the various curvatures at a given point are forced to add to the same value that they would have in hyperbolic space; in the case of (a higher dimensional version of) ``Circle Limit IV'', imagine that one distorts the geometry so that it looks ``more spherical'' in some directions, but ``more hyperbolic'' in other directions, taking care that the distortions all sum to zero. (The term ``Einstein manifold'' arises, of course, from General Relativity theory, in which certain important spacetimes (\emph{though not all}) do satisfy the Einstein condition.)

Thus, if we replace hyperbolic space by an asymptotically hyperbolic Einstein manifold, we are essentially allowing greater generality but ``within reason'': the asymptotic region is not (much) changed\footnote{In particular, there is a well-defined (direction-independent) negative curvature there, and one can again define a parameter $\rho$ by equating this curvature to $-1/\rho^2$.}, and the distortions have to be done in such a way that ``on average'' they do not change the curvature of hyperbolic space itself.

It is possible to prove, with the addition of a technical condition\footnote{For the curious: the condition is that the boundary of the conformal compactification, with the natural induced conformal structure, should have a non-negative Yamabe invariant. All of the boundary manifolds considered here have, in fact, a vanishing Yamabe invariant.} (which we need not discuss here, because it is always satisfied throughout our discussion), the following result \cite{kn:lee,kn:wang,kn:wang2,kn:galloway}:

\emph{If an asymptotically hyperbolic space is an Einstein manifold, then, for every surface homologous to the boundary at infinity (that is, a surface like the one at $r = R$ considered above), the analogue of inequality (\ref{D}) is satisfied.}

Thus, we have a rather clear picture of the kinds of generalizations of hyperbolic space for which an isoperimetric inequality can be expected to exist: the Einstein manifolds.

Now we turn to what appears to be a radically different world.

\addtocounter{section}{1}
\section* {\large{\textsf{2. The AdS/CFT Correspondence}}}
A major field of research in contemporary fundamental physics is the \emph{AdS/CFT correspondence} put forward by Juan Maldacena \cite{kn:malda} in 1997. To state it, we first need to define AdS, or anti-de Sitter spacetime. This is a spacetime with a metric obtained from the one in equation (\ref{B}) above by means of an apparently minor change: just reverse the sign of the term involving $dt^2$:
\begin{equation}\label{E}
g\;=\;dr^2\;+\;e^{2r/\rho}\,\left(-\,dt^2\;+\;dx^2\;+\;dy^2\right).
\end{equation}
A manifold with such an object defined on it (a non-singular bilinear form of signature (-1,1,1,1)) is called a semi-Riemannian manifold; in General Relativity, such manifolds are used to represent spacetime, so that indeed $t$ now represents time. This is anti-de Sitter (AdS) spacetime: it is the semi-Riemannian analogue of hyperbolic space. Similarly, one has asymptotically AdS spacetimes, exactly analogous to the asymptotically hyperbolic spaces discussed earlier. AdS is an extremely strange spacetime, not at all like the one we inhabit, but its theoretical interest is very great.

Now let us consider what happens when $r$ becomes extremely large. Then the exponential factor ensures that the term $dr^2$ becomes increasingly negligible\footnote{Again, this can be formulated rigorously, but that need not detain us. I will dispense with such observations henceforth, particularly since we are now talking about physics.}, and so, up to an overall factor, the geometry ``at infinity'' is \emph{flat}: the metric is just $-\,dt^2\;+\;dx^2\;+\;dy^2$.
Now this geometry is just (the three-dimensional version of) the ordinary geometry in which most of physics outside General Relativity is done: it is \emph{Minkowski spacetime.} It is, for example, the geometry used by particle physicists when they discuss what happens in particle colliders, such as the celebrated one at CERN in Geneva. Particle theorists discuss the interactions of quarks and other particles using \emph{quantum field theories}, the quantum mechanical generalizations of electromagnetic (and other) fields. A particular class of such quantum field theories is given the name of \emph{Conformal Field Theories} (CFTs): here the word ``conformal'' means that these theories are indifferent to the overall metric factors like the one we discarded above.

Maldacena's epochal claim, based on certain considerations in string theory (and subsequently buttressed by many theoretical tests) is, in broad terms, this: the physics of \emph{three-dimensional} CFTs defined on the flat boundary of an asymptotically AdS spacetime \emph{completely} captures the physics of any system in that \emph{four-dimensional} spacetime.

To understand how extraordinary this claim is, consider how broad is the class of asymptotically AdS spacetimes: it includes, for example, the AdS analogue of a massive, electrically and magnetically charged black hole endowed with angular momentum. An example \cite{kn:shear} of a metric for such a geometry is as follows:
\be
\label{F}
g(\ell {\rm dyKMV}_0)=-\frac{\Delta_r\Delta_\psi\rho^2}{\Sigma^2}\,\dif t^2+\frac{\rho^2\dif r^2}{\Delta_r}+\frac{\rho^2\dif\psi^2}{\Delta_\psi}+\frac{\Sigma^2}{\rho^2}\left[\omega\dif t-\dif\zeta\right]^2,
\ee
where the coordinates are $(t, r, \psi, \zeta)$ and where
\ba
\label{G}
\rho^2&=&r^2+(\ell-a\psi)^2\cr
\Delta_r&=&\frac{(r^2+\ell^2)^2}{L^2}-8\pi M^*r+a^2+4\pi \left[Q^{*2}+P^{*2}\right]\cr
\Delta_\psi&=&1+\frac{\psi^2}{L^2}(2\ell-a\psi)^2\cr
\Sigma^2&=&(r^2+\ell^2)^2\Delta_\psi-\psi^2(2\ell-a\psi)^2\Delta_r\cr
\omega&=&\frac{\Delta_r\psi(2\ell-a\psi)-a(r^2+\ell^2)\Delta_\psi}{\Sigma^2};
\ea
here $M^*, a, \ell, Q^*, P^*$ are certain constant parameters describing the physics of the black hole. According to Maldacena, such an enormously complex, fully four-dimensional object can be completely understood by studying some relatively simple conformal field theory defined infinitely far away on a simple spacetime of only three dimensions. Because holograms are two-dimensional objects which appear to be three-dimensional, this astonishing claim sometimes goes by the name of \emph{holography}.

How can a four-dimensional system be fully equivalent to a three-dimensional system? The only possible answer is that the four-dimensional system must be constrained extremely tightly, so that it does not contain as much information as it seems to do. In other words, the physics of asymptotically AdS systems must be governed by extremely restrictive conditions, or new ``laws of Nature'', beyond those we know. Most of these are unknown, and much of contemporary theoretical physics research amounts to a search for these conditions.

The AdS/CFT correspondence essentially states that \emph{by knowing something about the physics of the boundary of an AdS-like spacetime, one immediately knows something about the physics of what is happening deep inside}. This phrase should look familiar ---$\,$ it is precisely how we described isoperimetric inequalities in the preceding section. So it was remarkable and beautiful when, in 2014, Ferrari and Rovai \cite{kn:ferrari1,kn:ferrari2,kn:ferrari3,kn:ferrari4} (see also \cite{kn:84}) showed that \emph{one of the internal consistency conditions for the AdS/CFT correspondence is precisely the inequality} (\ref{D}). To put it as provocatively as possible: the AdS/CFT correspondence works \emph{because} the isoperimetric inequality is satisfied.

But \emph{is} it satisfied? Well, we saw above that it certainly will be, if the four-dimensional ``bulk'' (in physics parlance) is an Einstein manifold. Now the Einstein condition is very popular among differential geometers: it has turned out to be a very natural generalization of the condition of being a space of constant curvature, and there is a large literature on it, including a thick and well-regarded tome authored by A.L. Besse \cite{kn:besse}, a relative of N. Bourbaki. Ironically, however, there are important \emph{physics} applications in which the Einstein condition must be abandoned. Let us consider one of these.

\addtocounter{section}{1}
\section* {\large{\textsf{3. The Holography of Magnetic Fields}}}
In recent years, it has come to be realised that incredibly intense magnetic fields can arise in realistic physical systems: for example, a certain kind of neutron star (called a ``magnetar'') can have magnetic fields in the region of $10^{10}$ tesla (the standard unit for magnetic fields); by comparison, the reader's brain, as a result of reading this work, generates a magnetic field of around $10^{-12}$ tesla, while the strongest magnets that can be bought on the internet (neodymium-based ``rare-earth'' N52 magnets), which are indeed astonishingly strong, generate a field of around 1 tesla. The magnetic field of a magnetar would be lethal to humans at a distance of order 1000 kilometres, by disrupting the chemistry of the body. Yet even magnetar fields are dwarfed by fields which can be produced (briefly) in the laboratory.

Particle colliders, such as the famous LHC in Geneva, can be used to collide not just particles but entire atomic nuclei. When this is done, the nuclei often collide ``off-centre'', and the resulting swirling motion gives rise to fantastically large magnetic fields, perhaps as high as $10^{15}$ tesla. Matter under these extreme conditions is not well-understood, and so holographic methods (among others, of course) are employed: the magnetic field is regarded as residing on the boundary of an asymptotically AdS spacetime, as described in the previous section.

The matter formed in these collisions is extremely hot (with temperatures running into trillions of degrees), so one needs a holographic description which takes this, too, into account. This is done by considering an asymptotically AdS black hole: as is well known, black holes radiate thermally (they give off Hawking radiation), and this temperature can be adjusted to the desired value by adjusting the black hole parameters.

Up to this point, the ``bulk'' spacetime is an Einstein space, so the isoperimetric inequality holds. However, in order to obtain a magnetic field at infinity, one needs to put a magnetic charge on the black hole. The key point is this: the magnetic charge naturally gives rise to a magnetic field (in the ``bulk'' as well as on the boundary). In accordance with General Relativity theory, the presence of any form of matter in the spacetime alters its geometry, and one can show that, when the matter in question is a magnetic field, \emph{it then ceases to be an Einstein manifold}. It is therefore no longer clear that the isoperimetric inequality holds, and so the internal consistency of the theory is problematic.

The present author has shown \cite{kn:82} that the isoperimetric inequality continues to hold even in the presence of a magnetic field, \emph{provided} that this field is not too strong. The question, of course, is to define ``too strong''.

In \cite{kn:82} we proceeded as follows. If one wishes to increase the magnetic field on the boundary, one needs to change the magnetic charge\footnote{The reader may be aware that magnetic charge has never been observed. This is not relevant here, as we take the view that the ``bulk'' is a mathematical construct, not a ``real'' spacetime.} on the black hole in the bulk. When this is done, it affects, as we mentioned above, the geometry of the bulk spacetime. In particular, it produces a subtle change in the relative magnitudes of areas and volumes, and this is why it can lead to a violation of the isoperimetric inequality. However, changing the magnetic charge \emph{also} has an effect on the Hawking temperature of the black hole, and this gives us a way of specifying how strong the magnetic field should be in order to qualify as ``too strong'': one has to compare it with the temperature (actually, with the square of the temperature).

The final result is remarkably simple: one finds that the isoperimetric identity will be violated ---$\,$ that is, the holographic principle will be violated ---$\,$ if the magnetic field $B$ and the temperature $T$ fail to satisfy the following inequality (where we are using so-called ``natural'' units):
\begin{equation}\label{H}
B\;\leq \;2\pi^{3/2}T^2\;\approx \; 11.14 \times T^2.
\end{equation}
One might say that this \emph{is} the physics version of the isoperimetric identity, at least in this particular application.

Is this inequality actually satisfied, in experiments? The RHIC experiment \cite{kn:STAR}, at the Brookhaven National Laboratory in the United States, collides gold nuclei at energies sufficient to produce a system with extremely high temperatures and, in off-centre collisions, extremely high magnetic fields. The reported values of $B$ and $T^2$ from that experiment are roughly similar, so the inequality (\ref{H}) is indeed satisfied. However, in view of the huge quantities involved, and the inevitable experimental uncertainties, the factor of $\approx 10$ by which the right side exceeds the left is somewhat uncomfortably small.

The ALICE experiment \cite{kn:ALICE} at CERN in Geneva collides lead nuclei at higher energies, but the temperature of the system is only about 40$\%$ higher than at the RHIC, while the magnetic fields might be as much as 15 times as large; so the ``isoperimetric inequality'' still holds, but by a still smaller margin. It is likely that the inequality will be violated at some future facility, perhaps one which might be built in future in China \cite{kn:yau}. If this happens, and if AdS/CFT is correct, then we must be missing something: there must be some physical effect which we are neglecting, and which changes the bulk geometry in such a way that the isoperimetric inequality is protected. What that effect might be is a subject of current research\footnote{Note added: for recent progress on this question, see \cite{kn:88}.}.

\addtocounter{section}{1}
\section* {\large{\textsf{4. Conclusion}}}
The fact that highly sophisticated pure mathematics can be useful in physics has been celebrated many times. What is often not appreciated is the unexpected way in which this often happens. We began by considering a very ancient and beautiful pure mathematics problem, the study of the classical isoperimetric inequality. This inequality has analogues in many much more complicated spaces than the flat plane; and it turned out that one of these analogues places an entirely unexpected constraint on the behaviour of matter under the most extreme conditions we can test experimentally. We concluded by showing that the most advanced experiments on heavy ions currently under way actually push an isoperimetric inequality to its limits. In this sense, CERN is actually performing ``mathematical experiments''!

\addtocounter{section}{1}
\section*{\large{\textsf{Acknowledgement}}}
The author is grateful to Cate McInnes for useful discussions.

\end{document}